\documentclass[conference]{IEEEtran}

\IEEEoverridecommandlockouts
\usepackage{cite}
\usepackage{amsmath,amssymb,amsfonts}

\usepackage{graphicx}
\usepackage{textcomp}
\usepackage{xcolor}
\usepackage[ruled,linesnumbered]{algorithm2e}
\usepackage{wrapfig}
\usepackage{epsfig,endnotes,url,color,subfig}
\usepackage{booktabs}
\usepackage{enumitem}
\usepackage{hyperref}
\usepackage{float}

\usepackage{placeins}

\newcommand{\myparagraph}[1]{\smallskip \noindent{\bf {#1}.}}

\def\BibTeX{{\rm B\kern-.05em{\sc i\kern-.025em b}\kern-.08em
    T\kern-.1667em\lower.7ex\hbox{E}\kaern-.125emX}}

\begin{document}

\title{Entangled Futures: Quantum Computing's Potential and Challenges in Computer Science}


 \author{
     Devki Nandan Jha\IEEEauthorrefmark{1}, Jonte R. Hance\IEEEauthorrefmark{1}, Tomasz Szydlo\IEEEauthorrefmark{1}, Nick Chancellor\IEEEauthorrefmark{1}, Omer F. Rana\IEEEauthorrefmark{2},  Rajiv Ranjan\IEEEauthorrefmark{1} \\
     \IEEEauthorblockA{\IEEEauthorrefmark{1} Newcastle University, Newcastle upon Tyne, UK}
     \IEEEauthorblockA{\IEEEauthorrefmark{2} Cardiff University, Cardiff, UK}
}

\maketitle

\begin{abstract}
Quantum computing represents a fundamental shift in computation, promising to solve problems currently intractable for classical computers. However, misconceptions persist; quantum computers are not universally faster, nor do they replace classical systems. Current Noisy Intermediate-Scale Quantum (NISQ) devices face significant challenges, including decoherence, high error rates, and limited scalability. Additionally, leveraging quantum systems requires rethinking traditional computing principles, adopting different programming models, and addressing the complexities of integrating quantum and classical workflows, making their practical use in computer science both demanding and non-trivial. This article provides a system-level perspective on quantum computing for computer scientists, software engineers, and IT professionals, examining its current capabilities, key architectural and software challenges, and potential paths toward quantum advantage in real-world applications within hybrid quantum–classical environments.
\end{abstract}


\section{Introduction}
Quantum computing has redefined computational paradigms by leveraging the principles of quantum mechanics to address problems beyond the reach of classical systems. It encompasses multiple implementations, each based on distinct physical phenomena for processing quantum information \cite{lapierre2021introduction,de2021materials}. Among these, qubit-based quantum computers, often realised through solid-state technologies like superconducting circuits, trapped ions and photons, are the most extensively explored for general-purpose computing. These systems function as co-processors, relying on classical computers for critical tasks such as initialisation, control, error correction, and measurement. This hybrid quantum-classical computing is essential for practical applications, with classical systems handling pre- and post-processing while quantum processors focus on specialised computational tasks. The rise of cloud-based quantum platforms, offered by providers such as IBM, Google, D-Wave, and Rigetti, has further accelerated accessibility, allowing researchers to execute quantum programs through classical interfaces without direct hardware interaction.

This article approaches quantum computing from a computer science and systems perspective, focusing on integration challenges, software infrastructure, hybrid orchestration, and architectural considerations relevant to IT professionals. The current quantum computing architecture consists of the following components.

\begin{figure*}[t!]
    \centering
    \includegraphics[width=0.8\linewidth]{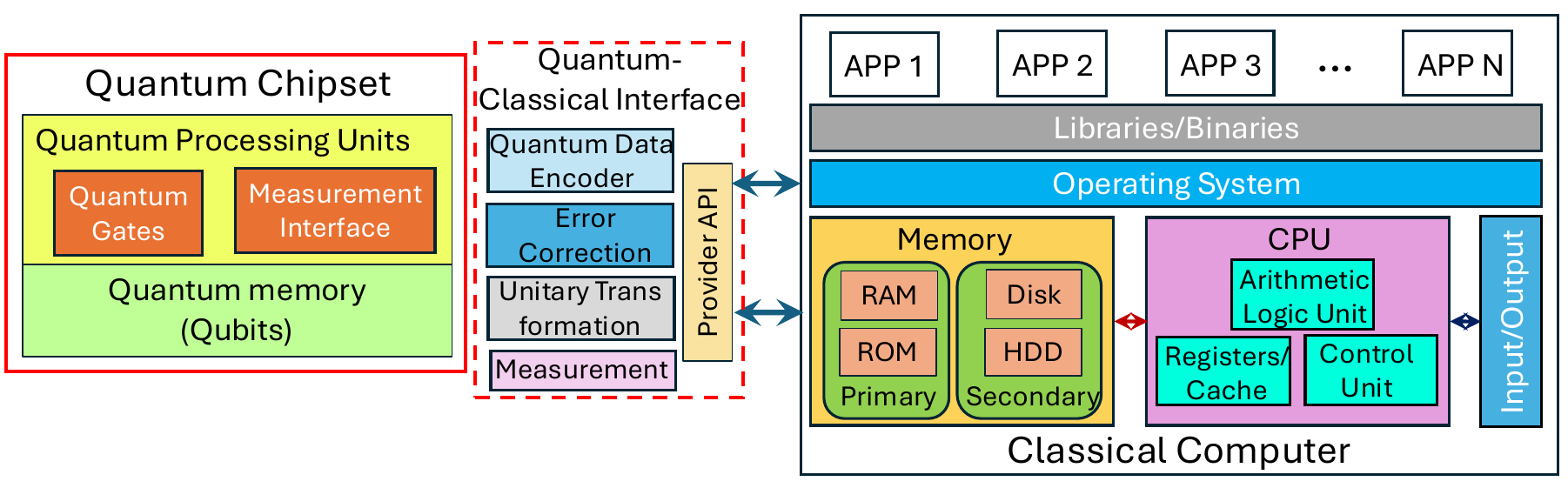}
    \caption{Schematic block diagram of Quantum Computing}
    \label{fig:QC_architecture}
    \vspace{-1em}
\end{figure*}

\myparagraph{Qubits}
Qubits or quantum bits are the basic units of quantum information in a quantum computer. Unlike classical bits, which exist in a state of either $0$ or $1$, qubits can be described by a point on the Bloch sphere: a geometric representation of a two-level quantum state as a point on a unit sphere, representing a continuum of possible states, which enables a much richer information structure than in classical computation \cite{lapierre2021introduction}.
Qubits can be realised using various physical systems, such as superconducting circuits, trapped ions, photons and quantum dots, each with its advantages and challenges \cite{de2021materials}.

\myparagraph{Quantum Processing Units (QPUs)} 
At the heart of a quantum computer is the Quantum Processing Unit (QPU), the computational engine that plays a role similar to the CPU in classical systems. There are several implementations of QPUs, including gate-based systems, quantum annealers, analogue quantum simulators, and measurement-based architectures, among others. While each approach has its strengths and weaknesses, gate-based QPUs are currently the most prominent due to their flexibility and alignment with general-purpose quantum algorithms. A gate-based QPU typically comprises physical qubits, quantum logic gates implemented via precise control pulses, and the classical control electronics necessary to orchestrate these operations \cite{bertels2020quantum}. 

Qubits are manipulated using quantum gates. Common gates include the Hadamard gate, which creates superpositions; the Pauli-X gate, which performs a bit-flip; the Pauli-Z gate, which applies a phase-flip; and the CNOT gate, which entangles qubits. These gates enable the powerful behaviours that distinguish quantum computation from classical computation. Quantum Control Systems apply signals such as microwave pulses (for superconducting qubits) or laser pulses (for trapped ions) to manipulate qubits. This system ensures that gate operations occur with exact timing and amplitude, maintaining the fidelity of quantum operations.

\myparagraph{Quantum Memory}
Quantum Memory is another essential component, designed to store quantum states. Technologies such as superconducting resonators, trapped ions, and atomic ensembles are being explored to provide quantum memory capabilities. Unlike classical memory, which retains stable bits (0s and 1s) indefinitely, quantum memory needs to maintain coherence i.e., remain in the current state for extended durations without collapsing due to environmental interference \cite{nico2024can}. 
However, current implementations have extremely short storage times, typically ranging from microseconds to seconds, 
depending on the physical system used, but maintaining long coherence times while ensuring scalability remains a fundamental challenge. Additionally, quantum error correction codes (QECC) present a way to 
preserve quantum states, but implementing them requires redundant qubits, adding to the complexity \cite{Roffe2019ErrorCorrect}. 



\myparagraph{Quantum Network}
A Quantum Network extends the capabilities of quantum computing and communication by enabling the secure transmission and distribution of quantum information across multiple nodes over long distances. 
These networks are foundational for the realisation of a quantum internet, which promises applications such as distributed quantum computing 
and long-range entanglement-based protocols. However, maintaining quantum coherence over long distances is a significant challenge due to the potential loss of quantum states and environmental noise during transmission. 

To overcome these limitations, quantum repeaters, intermediate devices designed to extend the reach of quantum communication by creating and relaying quantum states between distant nodes, are being developed. Quantum repeaters do not amplify quantum signals like classical repeaters do, since quantum information cannot be copied (as per the no-cloning theorem) \cite{benchasattabuse2024engineering}. Instead, they rely on advanced protocols such as entanglement swapping and quantum teleportation to relay quantum states over long distances. 


\myparagraph{Quantum-Classical Interface}
The \emph{Quantum-Classical Interface} plays a pivotal role in leveraging quantum computers for executing applications by facilitating seamless input/output interaction between classical and quantum computational paradigms. 
Quantum data encoders are responsible for translating classical input data into quantum states. Once the data is encoded, \emph{unitary transformations} are applied, representing the quantum computation. These transformations are implemented through a series of quantum gates. The quantum system is then \emph{measured} in a specific way, chosen such that the measurement results can be used to derive the desired solution to the computation.

To facilitate access to quantum computing, providers like IBM, Google, and Rigetti offer APIs that act as gateways for users to submit quantum programs to their cloud-based quantum computers \cite{laouid2025quantum}. These APIs abstract the complexities of hardware-specific operations and allow developers to interact with quantum devices through high-level programming frameworks such as Qiskit, Cirq, or Forest. This integration simplifies the deployment of quantum applications, enabling users to define circuits, manage jobs, and retrieve results without directly handling low-level hardware.

A schematic diagram showing the basic components and interactions in a quantum-classical system is shown in Figure \ref{fig:QC_architecture}.

\section{Open issues in utilising Quantum Computers in Computer Science}
The use of quantum computers in computer science presents numerous opportunities but also exposes a range of open issues that require further exploration. The following subsections discuss each in detail, highlighting gaps in current approaches and outlining potential research directions.

\subsection{Quantum algorithm design}
Quantum computing approaches fall broadly into three categories \emph{(i)  true quantum computation}, where algorithms run directly on a QPU and rely on quantum interference and entanglement for their advantage (e.g., Shor's and Grover's algorithms); \emph{(ii) hybrid quantum–classical variational methods}, where a parameterised quantum circuit is evaluated on a QPU while a classical optimiser updates circuit parameters iteratively; and \emph{(iii) quantum-inspired classical algorithms}, such as tensor network methods, that simulate aspects of quantum behaviour on conventional hardware without requiring a QPU. 

Developing new quantum algorithms is a major challenge due to the limited number of established solutions. While algorithms such as Shor’s and Grover’s provide well-known theoretical advantages over classical counterparts, the practical quantum advantage of near-term hybrid approaches, most notably the Variational Quantum Eigensolver (VQE), remains an active area of research \cite{hull2024quantum}. VQE exemplifies the hybrid variational paradigm, where a parameterised quantum circuit (the ansatz) is executed on a QPU to estimate an observable, and the resulting measurement outcomes are fed into a classical optimiser that updates the circuit's parameters. This loop repeats until convergence. VQE is closely related to the Quantum Approximate Optimisation Algorithm (QAOA), which applies the same variational principle to combinatorial problems rather than eigenvalue estimation \cite{cerezo2021variational}. 
Although VQE is a cornerstone of the NISQ era, its ability to consistently outperform state-of-the-art classical methods has yet to be demonstrated. The aim to leverage quantum speedups while maintaining classical flexibility faces unresolved challenges in optimisation and resource management \cite{Callison2022Hybrid}.  Other significant challenges include barren plateaus in optimisation landscapes, where gradients of the cost function vanish exponentially with circuit width and depth, making parameter training infeasible for large circuits \cite{larocca2025barren}. This phenomenon arises from a fundamental stress between circuit expressibility, the ability to explore a wide range of quantum states, and trainability, where highly expressive circuits tend to produce gradients that are indistinguishable from noise. Sensitivity to hardware noise and considerable classical computational overhead required in the hybrid optimisation loop further complicate practical deployment.


Beyond hybrid variational algorithms, quantum machine learning (QML) has emerged as a prominent research direction at the interface of quantum computing and classical machine learning \cite{biamonte2017quantum, schuld2021machine}. QML seeks not only to leverage quantum resources for enhanced learning and optimisation tasks, but also to improve the performance and scalability of quantum computing technologies themselves. By integrating parameterised quantum circuits into classical learning workflows, QML frameworks provide flexible tools for both data-driven applications and quantum system design \cite{melnikov2023quantum}. Recent advances demonstrate that QML can support the development and operation of quantum hardware through tasks such as automated device calibration, noise characterisation, error mitigation, and adaptive control \cite{cerezo2022challenges}. 
Despite these promising developments, substantial further research is required to rigorously assess scalability, robustness, and practical advantage under realistic NISQ settings.

The challenges in algorithm design emphasise the need for innovative research in quantum-specific and hybrid quantum-classical algorithms.  
Researchers must develop methods to translate practical problems into quantum circuits while addressing issues like noise resilience and hardware constraints. As the field progresses, it is crucial to identify classes of problems where quantum computers can provide substantial advantages and create algorithms tailored to those domains. 

\subsection{Application specification and mapping} 
Quantum computing introduces distinct challenges in specifying and mapping traditional computer science applications into quantum-compatible forms. Unlike classical systems, where application specification involves defining deterministic data structures and control flows, quantum computing requires an entirely new paradigm. Classical constructs like arrays, lists, and trees cannot be directly represented in quantum systems, as qubits can exist in superposition states, meaning measurement output can be probabilistic \cite{ali2022software}. This fundamental difference demands a rethinking of how information is structured and processed. Developers must frame applications in terms of quantum states and transformations while adhering to the constraints of quantum mechanics, such as no-cloning and state collapse upon measurement. These constraints make it challenging to design quantum-friendly specifications that accurately represent classical applications while leveraging quantum computing's unique advantages.

The process of translating an application into a format usable by a quantum computer is fundamentally tied to its logical representation in terms of quantum operations. Specifying an application involves defining how input data is encoded into quantum states, setting up state preparation protocols, and outlining the transformations needed to process this data.
This representation must account for the non-deterministic outcomes of quantum measurements. Developers must carefully design the flow of computation to ensure that intermediate quantum states are preserved until measurement and that the results can be meaningfully interpreted in the context of the original problem. This task is further complicated by the need to optimise quantum resource usage, such as minimising qubit counts, reducing circuit depth, and addressing hardware-specific limitations such as qubit connectivity and gate fidelity \cite{murillo2025quantum}.

\subsection{Lack of standardised compiler/transpiler} 
One critical challenge in quantum computing is the lack of standardised compilers and methodologies for translating classical problem definitions into quantum-compatible applications. Tools like Qiskit (IBM), Cirq (Google), Pennylane (Xanadu), and Tket (Cambridge Quantum) offer abstraction layers for circuit design, simulation, and execution. 
While they simplify quantum programming by abstracting low-level gate operations and hardware constraints, their tight coupling to specific platforms limits portability and interoperability. A circuit compiled for IBM’s quantum devices, for instance, may require re-optimisation or restructuring to run efficiently on Google’s hardware due to differences in gate sets and qubit topologies. This fragmentation slows application development and poses a barrier to scalability.

\begin{table}[]
\centering
\caption{Comparison of Quantum Programming tools}
\label{tab:tools}
\resizebox{\columnwidth}{!}{%
\begin{tabular}{|l|l|l|l|l|}
\hline
\textbf{Tools} & \textbf{Developer} & \textbf{Hardware} & \textbf{Features} & \textbf{Maturity} \\ \hline
Qiskit & IBM & \begin{tabular}[c]{@{}l@{}}IBM QPUs,\\ simulators\end{tabular} & \begin{tabular}[c]{@{}l@{}}Circuit model,\\ transpiler, runtime\\ primitives\end{tabular} & \begin{tabular}[c]{@{}l@{}}Very mature, large \\ eco-system, prod-\\ uction use\end{tabular} \\ \hline
Cirq & Google & \begin{tabular}[c]{@{}l@{}}Google QPU,\\ simulators\end{tabular} & \begin{tabular}[c]{@{}l@{}}NISQ circuits, \\ H/W-native gates\end{tabular} & \begin{tabular}[c]{@{}l@{}}Mature, Research-\\ focused, stable\end{tabular} \\ \hline
Q\# & Microsoft & \begin{tabular}[c]{@{}l@{}}Azure QPU,\\ simulators\end{tabular} & \begin{tabular}[c]{@{}l@{}}High-level language,\\ resource estimation\end{tabular} & \begin{tabular}[c]{@{}l@{}}Mature, enterprise \\ \& academic use\end{tabular} \\ \hline
pyQuil & Rigetti & Rigetti QPUs & Hybrid execution & Developing \\ \hline
\begin{tabular}[c]{@{}l@{}}Braket\\ SDK\end{tabular} & Amazon & \begin{tabular}[c]{@{}l@{}}AWS Braket,\\ simulators\end{tabular} & Multi-H/W & Developing \\ \hline
Tket & Quantinuum & H/W-agnostic & \begin{tabular}[c]{@{}l@{}}Advanced compila-\\ tion, optimisation\end{tabular} & Mature \\ \hline
PennyLane & Xanadu & H/W-agnostic & \begin{tabular}[c]{@{}l@{}}Hybrid QML,\\ ML-integration\end{tabular} & \begin{tabular}[c]{@{}l@{}}Growing, strong in \\ QML research\end{tabular} \\ \hline
\begin{tabular}[c]{@{}l@{}}QCGPU/\\ CUDA\\ Quantum\end{tabular} & NVIDIA & GPU + QPU & \begin{tabular}[c]{@{}l@{}}GPU Acceleration,\\ hybrid workflows\end{tabular} & Emerging \\ \hline
\end{tabular}%
}
\end{table}

The need for hardware-agnostic compilers and standardised intermediate representations is increasingly urgent. Emerging tools like OpenQASM 3.0 aim to provide cross-platform circuit representations, but they remain in early development \cite{kim2024enabling}. Multi-backend platforms like Amazon Braket offer unified interfaces for multiple hardware providers, helping to ease integration, but still rely on backend-specific compilation steps. Table \ref{tab:tools} gives a comparison of existing quantum programming tools available. Without a universal compiler that can optimise circuits across diverse hardware, truly portable and scalable quantum applications remain out of reach.


To address these challenges, research must focus on compilers that translate quantum programs into a hardware-independent format before applying platform-specific optimisations. These compilers should employ advanced techniques to minimise circuit depth, gate count, and error propagation for more efficient execution. Additionally, hybrid compilers that seamlessly combine classical pre/post-processing with quantum logic are essential for broader application use. Establishing standardised, flexible compiler infrastructures will be vital in bridging classical problem definitions with practical quantum execution, accelerating progress across the ecosystem. The compiler challenge is closely coupled to the scheduling problem discussed next, as compilation decisions directly affect coherence window usage.


\subsection{Scheduling and context switching in hybrid quantum-classical computing}
Scheduling and context switching in quantum-classical computing systems present fundamental challenges due to the contrasting nature of these paradigms. In classical systems, scheduling allocates resources like CPU cycles and memory, while context switching ensures multitasking by saving and restoring process states. However, quantum systems operate under strict coherence and error-rate constraints, making these tasks significantly more complex. 
As quantum coherence is extremely sensitive to environmental noise, tasks must complete within the coherence window; any pause or interruption risks decoherence and the irreversible loss of quantum state. This fragility demands careful resource management to minimise crosstalk and ensure optimal usage of limited quantum hardware \cite{li2024parallelizing}.

Progress has been made, platforms like IBM Quantum and Google’s Cirq implement queue-based systems that prioritise tasks based on complexity or user access. Compilation-aware schedulers in compilers like Qiskit and Tket reorder operations to minimise gate delays and idle time. Hybrid orchestration frameworks, such as Amazon Braket coordinate quantum and classical workflows for smoother handoffs. However, these systems are largely static, relying on predefined parameters and hardware assumptions that don’t account for real-time changes in qubit fidelity or noise. Runtime error mitigation techniques also increase scheduling complexity.

Complementing hardware execution and scheduling frameworks, classical simulators remain a foundational tool in the quantum software stack \cite{cicero2025simulation}. These simulators operate at multiple levels, from low‑level device or gate simulation to higher‑level algorithm emulation, and employ a variety of methods such as statevector propagation and hybrid approximation techniques to balance accuracy and performance. 
However, all classical simulators share a fundamental limitation: representing a general quantum state scales exponentially with the number of qubits, making faithful simulation of large or highly entangled systems infeasible in practice. Moreover, simulators often struggle to model realistic noise and open‑system effects at scale with high fidelity, which reduces their utility for evaluating error mitigation or control strategies that are critical in NISQ environment.

However, significant challenges remain. Quantum systems lack true context switching: ``quantum states can't be paused without collapsing", limiting multitasking. 
Dynamic scheduling algorithms are needed to adapt to real-time hardware variations. Future hybrid schedulers must support complex, interleaved workflows with minimal latency and efficient resource use. Multi-tenancy further complicates scheduling, requiring algorithms that balance fairness with efficiency. Solving these issues will be crucial for scaling quantum systems and integrating them into broader computing infrastructures.


\subsection{Communication in quantum environment} 
Effective communication between quantum systems and quantum-classical systems is essential for realising the full potential of quantum computing. Quantum channels like optical fibres and free-space communication are vulnerable to noise, decoherence, and photon loss, which degrade the fidelity of transmitted quantum states. The lack of scalable quantum repeaters further limits long-distance communication. While protocols like quantum teleportation and quantum key distribution (QKD) have shown promise, their widespread use is constrained by the absence of a robust quantum network infrastructure \cite{li2024survey}. Advancements are needed in high-fidelity repeaters, quantum error correction, and scalable networking to enable efficient interconnection between quantum processors.

In quantum-classical communication, the disparity between the probabilistic behaviour of quantum systems and the deterministic behaviour of classical systems introduces fundamental challenges. Measurement collapses quantum states into eigenstates with a definite value, making the bidirectional communication non-trivial. Moreover, quantum systems operate within brief coherence windows, while classical systems function continuously, causing synchronisation difficulties. Frameworks like Qiskit and Cirq offer partial solutions for managing execution and data exchange, but face limitations including high latency, bandwidth constraints, and lack of adaptive communication protocols.

Addressing these issues requires the development of faster analog-to-digital conversion methods, dynamic synchronisation protocols, and dedicated hardware for quantum-classical interfacing. Integrating quantum and classical networks into unified hybrid frameworks is also vital. This includes creating middleware for real-time workflow adaptation and error correction. Advancing the infrastructure to resemble classical internet systems will be key to scaling hybrid platforms. Research must focus on quantum routers, scalable network topologies, and multi-tenant communication protocols across quantum-quantum and quantum-classical contexts.

\subsection{Programming and Software Development}
One of the biggest challenges in quantum computing today is not just the fragile nature of qubits, it’s the software. Quantum programming remains surprisingly low-level, requiring developers to work with abstract quantum concepts. Even with modern frameworks like IBM’s Qiskit, Google’s Cirq, Microsoft’s Q\#, writing a quantum program still means building circuits gate by gate, often with careful attention to the peculiarities of specific hardware. This is in contrast to classical software development, where high-level languages, modular libraries, and reusable APIs have dramatically accelerated how we build and scale applications. As it stands, quantum programming is closer to circuit design than to application development.

This steep learning curve limits accessibility, particularly for developers outside of physics or quantum information science, thereby impeding innovation and collaboration across disciplines. Prototyping a new idea in quantum code can take days of manual circuit tuning and optimisation, where a classical developer might iterate through dozens of prototypes. There’s also very little in the way of standard libraries or pre-built modules for common tasks like state preparation or quantum arithmetic, meaning many teams are forced to reinvent solutions repeatedly.

Alternative programming models are emerging that suggest a way forward. It is important here to distinguish between two fundamentally different computational paradigms. Gate-model quantum computers (IBM, Google, IonQ) implement universal quantum computation through sequences of discrete unitary operations on qubits. They offer programmability and generality but remain hardware-constrained. On the other hand, Quantum annealers and Ising machines, such as D-Wave's systems, operate on an entirely different principle. They encode optimisation problems directly into physical energy landscapes and search for low-energy configurations corresponding to candidate solutions \cite{rajak2023quantum}.
While these approaches do not provide universal gate-based quantum computation, they have demonstrated practical utility in combinatorial optimisation and sampling tasks, often at scales currently challenging for gate model systems. D-Wave's QUBO and Ising interfaces allow users to frame problems in classical optimisation terms, abstracting the underlying physics. The quantum part is handled behind the scenes, letting developers focus on what they want to solve, not how to express it in quantum gates. While this is specific to one type of quantum hardware, it opens the door to broader ideas: what if gate-model quantum systems offered similar high-level wrappers for optimisation and simulation? 
Research into such abstractions, along with domain-specific languages, hybrid orchestration tools, and reusable quantum libraries, could dramatically improve accessibility and speed up innovation. 

\subsection{Monitoring and Debugging in Quantum Programs}
Debugging quantum programs presents a significant challenge due to the fragile nature of quantum information. Unlike classical computing, where developers can inspect variables and step through code, quantum computing lacks such transparency. Every measurement alters the system, making it impossible to observe a quantum program without affecting its outcome. Moreover, the probabilistic nature of quantum systems means that even correct programs may yield different results on each run. The complexity of software programming makes it even harder. Debugging, therefore, becomes more about inferring the behaviour of the quantum system rather than directly inspecting it. Tools such as Qiskit simulators and noise modelling help with simulating imperfections, but they can’t replicate the full unpredictability of real quantum hardware, especially as circuits grow in size and complexity \cite{yang2023survey}.

Researchers have begun exploring more advanced approaches. Quantum state tomography reconstructs quantum states through repeated measurements in multiple bases, but its measurement cost scales exponentially with system size \cite{anshu2024survey}. To mitigate this, techniques such as classical shadows compress measurement data into efficient statistical representations \cite{levy2024classical}. The recently introduced shadow overlap formalism further reduces measurement complexity for state certification \cite{huang2025certifying}. 
Other directions, such as weak measurements (which probe quantum states gently without fully collapsing them) and Symmetric Informationally Complete Positive Operator-Valued Measures (SIC-POVMs), which probe a quantum state simultaneously in all directions of the state space, extracting maximal information per measurement, offer ways to go beyond standard tomography through projective measurements in different orthogonal bases, though these techniques remain highly technical and not yet practical for routine debugging \cite{xing2025measuring}. Also, most of these methods are computationally expensive, hard to scale, and require deep domain expertise.

Future development in this area requires more scalable and accessible debugging tools, including hybrid debugging frameworks, where classical systems infer potential sources of error without disturbing the quantum state, and machine learning approaches for automated identification of circuit errors \cite{rishiwal2025new}. The security implications of such monitoring tools form part of the broader challenge discussed below.

\subsection{Security in Quantum Computing}
Security in quantum computing is a multi-dimensional challenge that covers computation, communication, and integration with classical systems. 
While quantum computers promise breakthroughs in problem-solving power, they also create new security dynamics that differ from classical systems. 
Today’s quantum systems are highly sensitive to noise, interference, and operational inconsistencies, requiring complex fault-tolerant architectures and QECCs to maintain stability. These solutions come at a steep cost as protecting a single logical qubit may require dozens or even hundreds of physical qubits, creating large attack surfaces and operational overhead.

While much of the quantum stack is new, many of its vulnerabilities are from classical systems, especially side-channel attacks, which typically target the classical control and readout layers rather than the quantum logic itself. The attackers might exploit electromagnetic leakage, power analysis, or timing variations in the classical electronics that interface with quantum hardware. Current defences, such as noise-resilient control protocols and shielding techniques, are still in early development and struggle to provide consistent protection across hardware platforms. More adaptive, low-overhead protection mechanisms that dynamically defend against evolving threats are needed. 
Additionally, quantum networking brings its own set of security issues. Protocols such as QKD promise theoretically unbreakable encryption, however, practical implementations are still vulnerable to real-world environments such as photon-number-splitting and detector blinding attacks, which target imperfections in physical devices. Moreover, integrating quantum networks with classical infrastructure compounds the issue, as classical systems do not inherently inherit quantum protections. 

Technologies such as the trusted platform module (TPM) and secure multiparty computation (SMC) can offer partial safeguards, but these are often insufficient for the diverse and dynamic nature of quantum-classical systems. More robust orchestration frameworks are needed to ensure data integrity and confidentiality as information moves between classical and quantum domains. Finally, while post-quantum cryptography (PQC) offers classical cryptographic methods designed to resist known quantum algorithms (e.g., Shor’s and Grover’s), it is important to note that their security is not absolute \cite{li2023post}. PQC is only as strong as our current understanding of quantum algorithm capabilities. Future quantum algorithms could potentially render current PQC methods vulnerable. Therefore, PQC should be viewed as a transitional safeguard rather than a definitive solution. A summary of issues discussed in this section is given in Table \ref{tab:issues}.

\begin{table}[]
\centering
\caption{Open Issues in Quantum Computing for Computer Science}
\label{tab:issues}
\resizebox{\columnwidth}{!}{%
\begin{tabular}{|c|c|c|c|}
\hline
\textbf{Challenge Area} & \textbf{Type} & \textbf{Tools/Approaches} & \textbf{Key Barriers} \\ \hline
\begin{tabular}[c]{@{}c@{}}Algorithm\\ Design\end{tabular} & \begin{tabular}[c]{@{}c@{}}Algorithmic/\\ Theoretical\end{tabular} & \begin{tabular}[c]{@{}c@{}}VQE, QAOA, Shor's,\\ Grover's,  QML\end{tabular} & \begin{tabular}[c]{@{}c@{}}Barren plateaus, noise,\\ limited provable advantage\end{tabular} \\ \hline
\begin{tabular}[c]{@{}c@{}}Application\\ Mapping\end{tabular} & S/W engg & \begin{tabular}[c]{@{}c@{}}Quantum S/W engg\\ toolchains\end{tabular} & \begin{tabular}[c]{@{}c@{}}No classical data structures,\\ probabilistic output\end{tabular} \\ \hline
\begin{tabular}[c]{@{}c@{}}Compiler/\\ Transpiler\end{tabular} & \begin{tabular}[c]{@{}c@{}}System/\\ Tooling\end{tabular} & \begin{tabular}[c]{@{}c@{}}Qiskit, Cirq, Tket, \\ Amazon Braket\end{tabular} & \begin{tabular}[c]{@{}c@{}}Hardware fragmentation,\\ no universal infrastructure\end{tabular} \\ \hline
\begin{tabular}[c]{@{}c@{}}Scheduling \& \\ Context \\ Switching\end{tabular} & \begin{tabular}[c]{@{}c@{}}Systems/\\ Runtime\end{tabular} & \begin{tabular}[c]{@{}c@{}}Quantum queue,\\ Qiskit schedulers, \\ Amazon Braket tools\end{tabular} & \begin{tabular}[c]{@{}c@{}}Coherence constraints,\\ no true context switching\end{tabular} \\ \hline
\begin{tabular}[c]{@{}c@{}}Quantum\\ Communication\end{tabular} & Networking & \begin{tabular}[c]{@{}c@{}}QKD, quantum\\ repeaters, teleportation\end{tabular} & \begin{tabular}[c]{@{}c@{}}Decoherence over distance,\\ no scalable repeaters\end{tabular} \\ \hline
\begin{tabular}[c]{@{}c@{}}Programming\\ \& S/W Dev\end{tabular} & \begin{tabular}[c]{@{}c@{}}Developer\\ Experience\end{tabular} & \begin{tabular}[c]{@{}c@{}}Qiskit, Cirq, Q\#, D-\\ Wave, QUBO/Ising\end{tabular} & \begin{tabular}[c]{@{}c@{}}Low-level abstraction,\\ steep learning curve\end{tabular} \\ \hline
\begin{tabular}[c]{@{}c@{}}Monitoring \&\\ Debugging\end{tabular} & \begin{tabular}[c]{@{}c@{}}Verification/\\ Testing\end{tabular} & \begin{tabular}[c]{@{}c@{}}Qiskit simulators, state\\ tomography, classical \\ shadows, SIC-POVMs\end{tabular} & \begin{tabular}[c]{@{}c@{}}No non-invasive inspection, \\ exponential measurement \\ cost\end{tabular} \\ \hline
Security & \begin{tabular}[c]{@{}c@{}}Trust/\\ Defence\end{tabular} & \begin{tabular}[c]{@{}c@{}}QKD, PQC, TPM,\\ SMC\end{tabular} & \begin{tabular}[c]{@{}c@{}}Classical attack surfaces,\\ hybrid system vulnerabilities\end{tabular} \\ \hline
\end{tabular}%
}
\end{table}

\section{Quantum computing vs Classical Computing}
Quantum computing is still in its infancy, far from replacing classical systems, but its true potential lies in hybrid quantum-classical integration. While existing NISQ processors have demonstrated advantages in many problems, their small qubit counts, high error rates, and noise limit their broad commercial utility today, although demonstrations of computational advantage continue to emerge in controlled settings. In contrast, classical computing has matured over decades, becoming the backbone of modern computing infrastructure. 

The near-term reality is a hybrid model rather than a self-contained quantum replacement. Application domains where hybrid quantum-classical workflows are actively being explored include computational chemistry (molecular energy simulation via VQE), combinatorial optimisation (logistics, scheduling via QAOA), and financial portfolio optimisation (sampling and risk analysis). In each case, quantum processors are envisioned as accelerators for specific computational kernels, while classical systems handle data management, orchestration, and result interpretation. It is important to distinguish between demonstrated advantage, achieved so far primarily in narrowly defined benchmark settings, and the broader speculative potential that remains subject to ongoing research. Cloud-based quantum services from IBM, Google, and Amazon already embed this hybrid model in practice: developers access QPUs as remote accelerators via classical APIs, with quantum jobs queued, scheduled, and post-processed through conventional infrastructure.

This hybrid approach is already shaping the future of computing, as researchers focus on refining quantum error correction, optimising quantum algorithms for NISQ devices, and building seamless quantum-classical software stacks. 
The future of computing is not quantum or classical alone, but a synergistic model where quantum co-processors unlock new possibilities, while classical computers provide the stability and efficiency needed for real-world applications.

\section*{Acknowledgment}
This study is partially supported by the UKRI National Edge AI Hub for Real Data: Edge Intelligence for Cyber-disturbances and Data Quality (EPSRC EP/Y028813/1), Royal Society Research Grant (RG/R1/251590), EPSRC Mathematical Sciences Small Grant (UKRI3647), EPSRC Quantum Technologies Career Acceleration Fellowship (UKRI1217), EPSRC QCI3 Quantum Computing Hub (EP/Z53318X/1), and EPSRC QEVEC (EP/W00772X/2).

\bibliographystyle{IEEEtran}
\bibliography{main}

\section*{Biography}

\textbf{Devki Nandan Jha} is currently a  Lecturer at Newcastle University, UK. He is also a visiting researcher at the Oxford e-Research Centre, University of Oxford. His research interests include cloud computing, quantum computing, embedded systems, trust and security, and machine learning.  Contact him at dev.jha@ncl.ac.uk.

\textbf{Jonte R. Hance} is Head of the Quantum Group, an EPSRC Quantum Technologies Career Acceleration Fellow, and a Senior Lecturer in Applied Quantum Foundations in the School of Computing at Newcastle University, UK.
 
Their research interests focus around trying to understand the paradoxical aspects of quantum mechanics, to leverage them to develop new and interesting technologies. Contact them at jonte.hance@ncl.ac.uk. 

\textbf{Tomasz Szydlo} is currently a Senior Lecturer at Newcastle University. He is also an Associate Professor at AGH University of Cracow, Poland. His research interests include IoT, Tiny ML, embedded systems and software integration. Contact him at tomasz.szydlo@ncl.ac.uk.

\textbf{Nick Chancellor} is a half-time lecturer at Newcastle University and works as a consultant in quantum computing (including as an advisor to Quantum Computing Inc.). His interests are hybrid quantum-classical algorithms, with a focus on quantum annealing and analog quantum computing techniques. He is noted for having written an early paper proposing a version of the reverse-annealing technique, which is now commonly used on D-Wave flux qubit devices. Contact him at nick.chancellor@ncl.ac.uk.

\textbf{Omer F. Rana} is a Professor of performance engineering with Cardiff University, Cardiff, United Kingdom. His current research interests include problem-solving environments for computational science and commercial computing, data analysis and management for large-scale computing, and scalability in high-performance agent systems. Contact him at ranaof@cardiff.ac.uk.

\textbf{Rajiv Ranjan} is a Professor of Computing Science in the School of Computing at Newcastle University, United Kingdom. He has a PhD (2009) from the Department of Computer Science and Software Engineering at the University of Melbourne. His research interests include distributed systems, Internet of Things and AI. Contact him at raj.ranjan@ncl.ac.uk.

\end{document}